\documentclass[12pt]{elsarticle}
\usepackage[textwidth=14.5cm]{geometry}
\usepackage{amsfonts}
\usepackage{url}
\usepackage{graphicx}
\usepackage{float}
\usepackage{xcolor}
\journal{Optics Communications}
\begin{document}
\begin{frontmatter}
\title{Ultra-broad band perfect absorption realized by phonon-photon resonance in periodic polar dielectric material based pyramid structure}

\author[1]{Kaidi Xu}
\author[1,2]{Gaige Zheng\corref{cor1}}
\cortext[cor1]{Corresponding author}
\ead{jsnanophotonics@yahoo.com}
\address[1]{The school of Physics and Optoelectronics Engineering, Nanjing University of Information Science \& Technology, Nanjing 210044, China}
\address[2]{iangsu Collaborative Innovation Center on Atmospheric Environment and Equipment Technology (CICAEET), Nanjing University of Information
Science \& Technology, Nanjing 210044, China}

\begin{keyword}
    \texttt{Perfect absorber, Surface phonon resonance, mid-infrared atmosphere window}
\end{keyword}

\begin{abstract}
    In this research, a mid-infrared wide-angle ultra-broadband perfect absorber which composed of pyramid grating structure has been comprehensively studied.
    The structure was operated in the reststrahlem band of SiC and with the presence of surface phonon resonance(SPhR), the perfect absorption was observed in the region between 10.25 and 10.85 $\mu m$. We explain the mechanism of this structure with the help of PLC circuit model due to the independence of magnetic polaritons.
    {Besides}, by studying the resonance behavior of different wavelength, we bridged the continuous perfect absorption band and the discret peak in 11.05 $\mu m$(emerge two close absorption band together) by modification of the geometry. The absorption band has been sufficiently broadened. More over, both 1-D and 2-D periodic structure has been considered and the response of different incident angles and polarized angles
    has been studied and a omnidirectional and polarization insensitive structure can be realized which may be a candidate of several sensor applications in meteorology.     
    The simulation was conducted by the Rigorous Coupled Wave Method(RCWA).
\end{abstract}

\end{frontmatter}

\section{Introduction}
Perfect absorber is a device which absorbs a certain range of light and converts it to some form of internal energy and then subsequently flows out to an external reservoir in the form of heat or current.
The first metamaterial perfect absorber {was} experimentally realized by Landy et al.\cite{landy2008perfect}
A new concept of perfect absorption by coherent illumination was proposed by Y.D.Chong et.al in 2010\cite{chong2010coherent}, according to whose theory perfect absorption was arise due to the time-reversal symmetry.
Also, the frequency and the relative phase was chosen to correspond a specific perfect absorption mode. This is also of great interest, which may be potentially useful in the various applications such as transducer, modulators and optical switches\cite{pu2012ultrathin}. Comparing with traditional incoherent-based metamaterial, the coherent control of absorption has additional tunability\cite{baranov2017coherent}. 
\par The perfect absorber can be divided into narrow-band and broad-band both of which has a great variety of applications at different frequencies (visible, near-infrared, mid-infrared, and THz).
Loads of studies of narrow-band have been reported like single-band\cite{wu2017ultra,yong2016narrow,li2016tunable}, dual-band{\cite{chen2012dual,zhang2015independently,yoo2013polarization}}, triple-band\cite{shen2011polarization,yoo2015triple}
and multi-band{\cite{sreekanth2016multiband,xia2017multi}} {which may be realized by dinstict resonators\cite{guan2020two,feng2014parallel}}. 
{In addition, in recent years, some of the most advanced 2D material that has been well developed in condensed matter physics such as graphene has attracted great interest. The corresponding studies of perfect absorption and transparency properties has also been proposed with different spatial arrangement of nano resonators\cite{xia2017multi,guan2020two,xia2020polarization}}
They have applications such as plasmonic sensor\cite{wu2017ultra,yong2016narrow,meng2014optimized}. Due to the narrow range of surface plasmon resonance (SPR) and SPhR, obtaining the broadband perfect absorber has always being a challenging task.
However, various ways has been developed to broaden the band of perfect absorber, by vertically stacking spacer {\cite{wang2012tunable,durmaz2019polarization,feng2014parallel,xia2020polarization}} and metamaterial based space-filling design -- using a certain algorithm to control the size and shape of resonators\cite{yu2019broadband,mcvay2005thin}.
\par In mid-infrared region, the phonon-photon resonance can be considered which is the counterpart of plasmonic resonance for polar dielectric structure\cite{wang2011phonon}. Unlike in metallic structure in which conductive are responsible for diamagnetism, the displacement current is in phonon-photon resonance structure. The suitable material is rather limited due to the requirement of large permittivity.
However, the {SiC} is a splendid choice which exhibit a phonon-polariton gap spectrum(reststrahlem band) between 10.3 to 12.6 $\mu m$\cite{catrysse2007near}.
This region is of high interest and closely related to the atmosphere window (8--14 $\mu m$). Therefore, it may be capable of application of atmosphere sensor.
\par In this paper, we bring out a {SiC-based} Pyramid-like grating structure which sufficiently increase the number of resonator and consequently broaden the widths of resonance peaks.
The RCWA simulation has demonstrated that the average absorption efficiency from 10.25 $\mu m$ to 10.85 $\mu m$ is over 99\% with the strongest absorption being 99.9997\%.
Different from previous reports, we have avoided vertically stacking the resonators or locate the resonators in a more complicated way but a complete geometry structure which may be more easily proceeded in the fabrication sense.
{On top of these, the polarization and omnidirectional properties has also been studied. Strong guided-moode resonance has been obseved  at different incident angle.}
\section{Structures and Design}
\begin{figure}[hbt!]
    \centering
    \includegraphics[width=30pc]{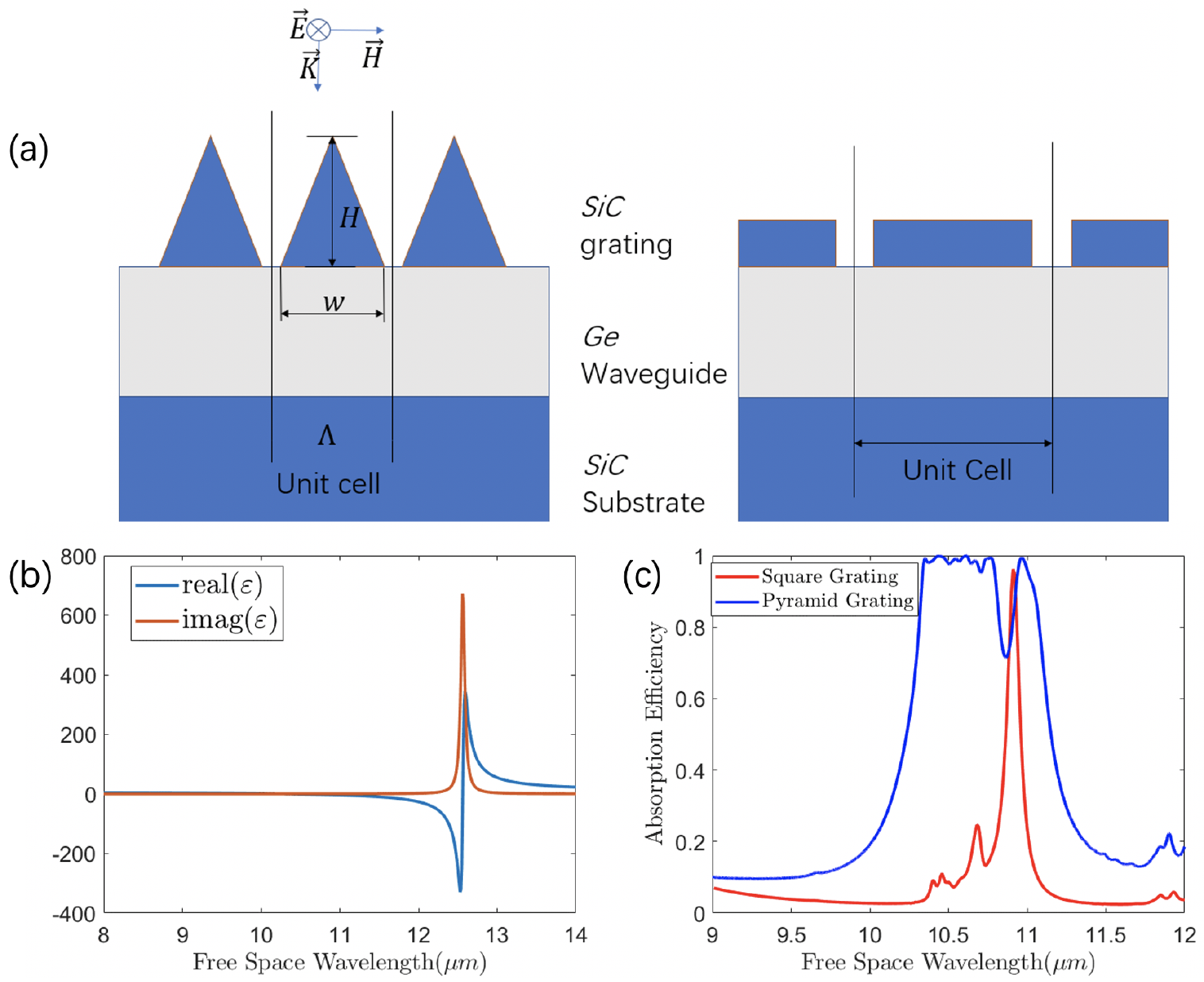}
    \caption{(a)The basic pyramid structure we used and the square grating structure as we have demonstrated. The real part and the imaginary part of permittivity is shown in (b). The absorption efficiency of each structure is illustrated in (c) with the blue line indicating the pyramid structure and red line indicating the square structure. As one can vividly see, the simulation result of RCWA manifests that we have realized the perfect absorption in the region of 10.25~$\mu m$ to 10.85~$\mu m$. and sufficiently larger than the square structure has obtained.}
    \label{fig1}
\end{figure}

The basic structure is shown in Fig.~\ref{fig1}(a) and a square structure has also been presented for an intuitive comparison.
The height of grating $H$, Filling length $W$, period $\Lambda$ is 5.4 $\mu m$, 4.8 $\mu m$, 5.2 $\mu m$ respectively.
A plane wave propagates along the z direction with the electric and magnetic fields polarized along the x and y directions, respectively.
For diffrent purpose, the grating can be designed as possessing translational symmetry in both x and y direction in which case, the structure may not be {sensitive to the polarization angles.}
\par In the absence of free carrier the frequency-dependent dielectric permittivity can be written in this form {by Lorentz-Drude model\cite{neuner2009critically,born2013principles}:}
\begin{equation}
    \varepsilon_{SiC}(\omega) = \varepsilon'_{SiC} + \varepsilon''_{SiC} = \varepsilon_{\infty}\frac{\omega^2-\omega_{LO}^2+i\gamma\omega}{\omega^2-\omega_{TO}^2+i\gamma\omega}.
\end{equation}
where $\omega$ is the wave number of incident light in free space and the longitudinal optical phonon frequency is $\omega_{LO} = 972~cm^{-1}$, the transverse optical phonon frequency is $\omega_{TO} = 796~cm^{-1}$, and $\varepsilon_{\infty}=6.5$ and
the damping rate arise from the vibriation of lattice and harmonicity is $\gamma =3.75~cm^{-1}$. From the discussion above, we know the Reststrahlem band corresponding to the region of $\omega_{TO}<\omega<\omega_{LO}$ and in the wavelength sense is $10.3~\mu m<\lambda<12.6~\mu m$. The real and imaginary part of permittivity is shown in {Fig.~\ref{fig1}(b).}
By applying the Cauchy dispersion model, one may get frequency-dependent refractive index. From the literature\cite{icenogle1976refractive}, the sellmeier coefficient is determined and {one writes the refractive index in this form:}
\begin{equation}
    n(\lambda) = \sqrt{9.28156+\frac{6.7288\lambda^2}{\lambda^2-0.44105}+\frac{0.21307\lambda^2}{\lambda^2-3870.1}}.
\end{equation}
In the region of our interest, the refractive index of {Ge} is approximately 4.005 with a vibration of the magnitude of $10^{-4}$. If one limited the order of fourier expansions in RCWA less than 200.
One may ignore the vibration and consider it as a constant.
\par From {Fig.~\ref{fig1}(c)}, it has shown that the perfect absorption band has been effectively broadened. In order to understand the mechanism of this structure, we present the magnetic field and displacement current density distribution (recall what we discuss above, the magnetic polariton was excited due to the displacement current in polar dielectric material).
What we expect to see is the strong magnetic field of different resonators corresponding to different absorption peak.
\begin{figure}[hbt!]
    \centering
    \includegraphics[width=30pc]{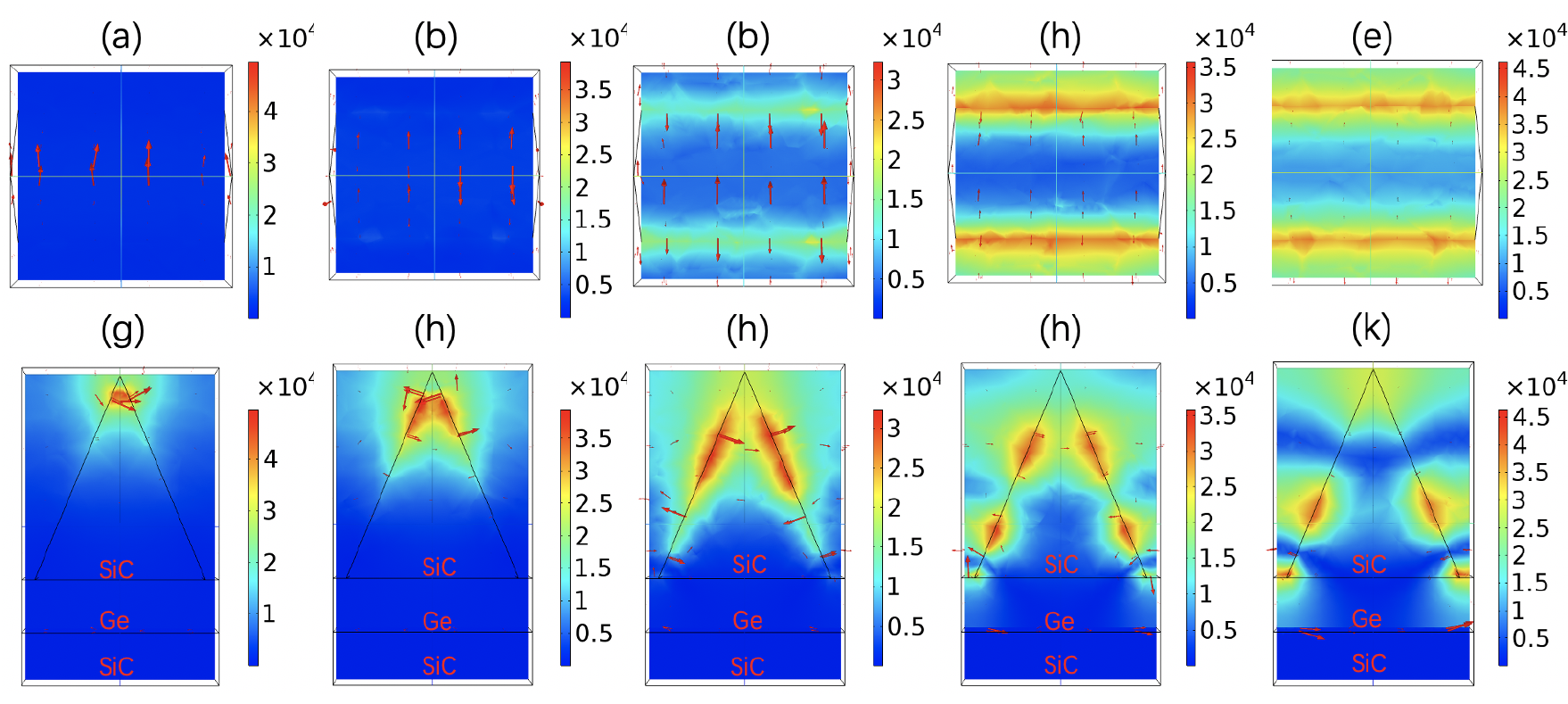}
    \caption{(a)(b)(c)(d)(e) indicate the magnetic field distribution in xy-plane when the wavelength of incident light is 10.3, 10.5, 10.6, 10.86(the dip) and 11 $\mu m$, respectively. And (f)(g)(h)(i)(j) indicate the corresponding magnetic field in xz-plane. The red arrows represent the displacement current flow.}
    \label{field_distribution}
\end{figure}
\par In Fig.~\ref{field_distribution}, one can vividly see that the different wavelength of perfect absorption corresponding to the different region of strong resonance in pyramid structure and the strong diplacement current flow occur in this very place.
As for the last peak(11 $\mu m$) after a dip in Fig.~\ref{fig1}, it has a more complex resonance behavior. 
\section{The polarization-insensitive design}
\par
\begin{figure}[hbt!]
    \centering
    \includegraphics[width=30pc]{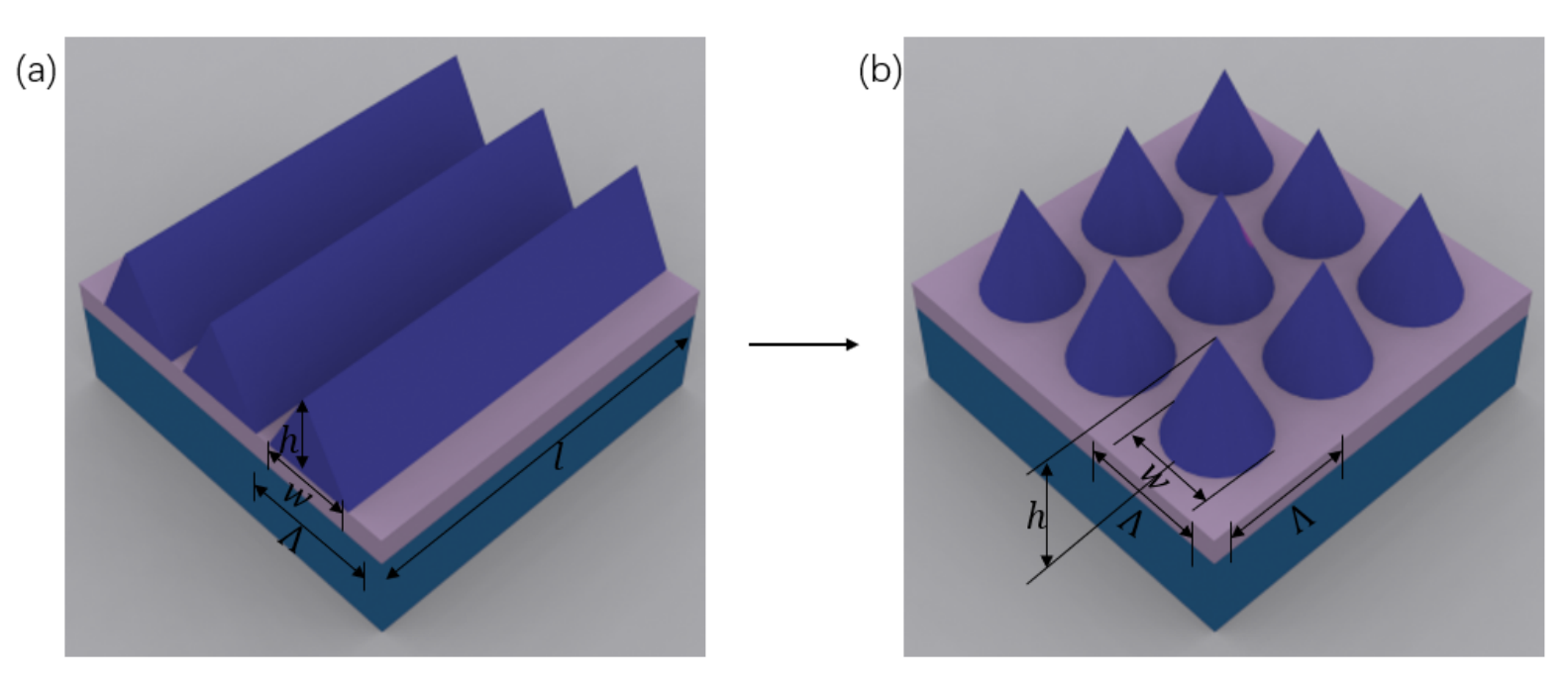}
    \caption{(a)The schematic of 1-D grating. (b)The schematic of 2-D grating}
    \label{2dgrating}
\end{figure}

 In order to obtain a {polarization-insensitive} perfect absorber, one may turn one's concern from 1D grating structure to 2D grating structure by applying the translational invariance in both x and y direction. In fact, a 2D structure possessing similar structure has also been studied and illustrated in Fig.~\ref{2dgrating}.  
\begin{figure}[hbt!]
    \centering
    \includegraphics[width=30pc]{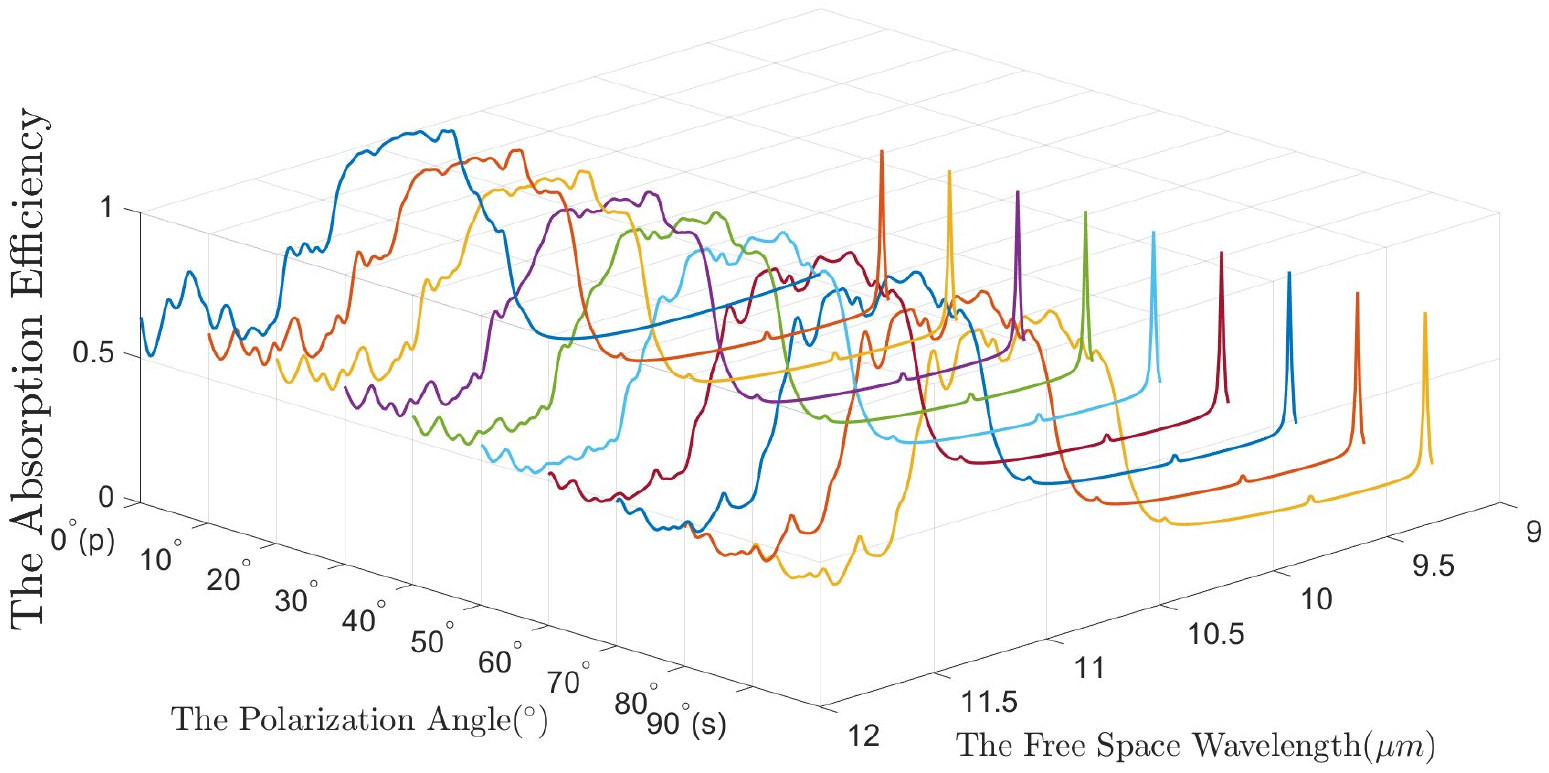}
    \caption{The spectral response of different polarization angle where x, y, z label indicate the polarization angle ($0^{\circ}$ and $ 90^{\circ} $ represent the p-polarized and s-polarized wave, respectively), the free space wavelength, the absorption efficiency, respectively}
    \label{polar}
\end{figure}
\par The spectral response of different polarization angles is shown in Fig.~\ref{polar}, from which one may find that the absorption spectrum has only a small shift among the regime of our interest with the change of polarization angle.
Also, the perfect absorption regime has a slight shift comparing with the corresponding 1-D structure.

\par
\begin{figure}[hbt!]
    \centering
    \includegraphics[width=30pc]{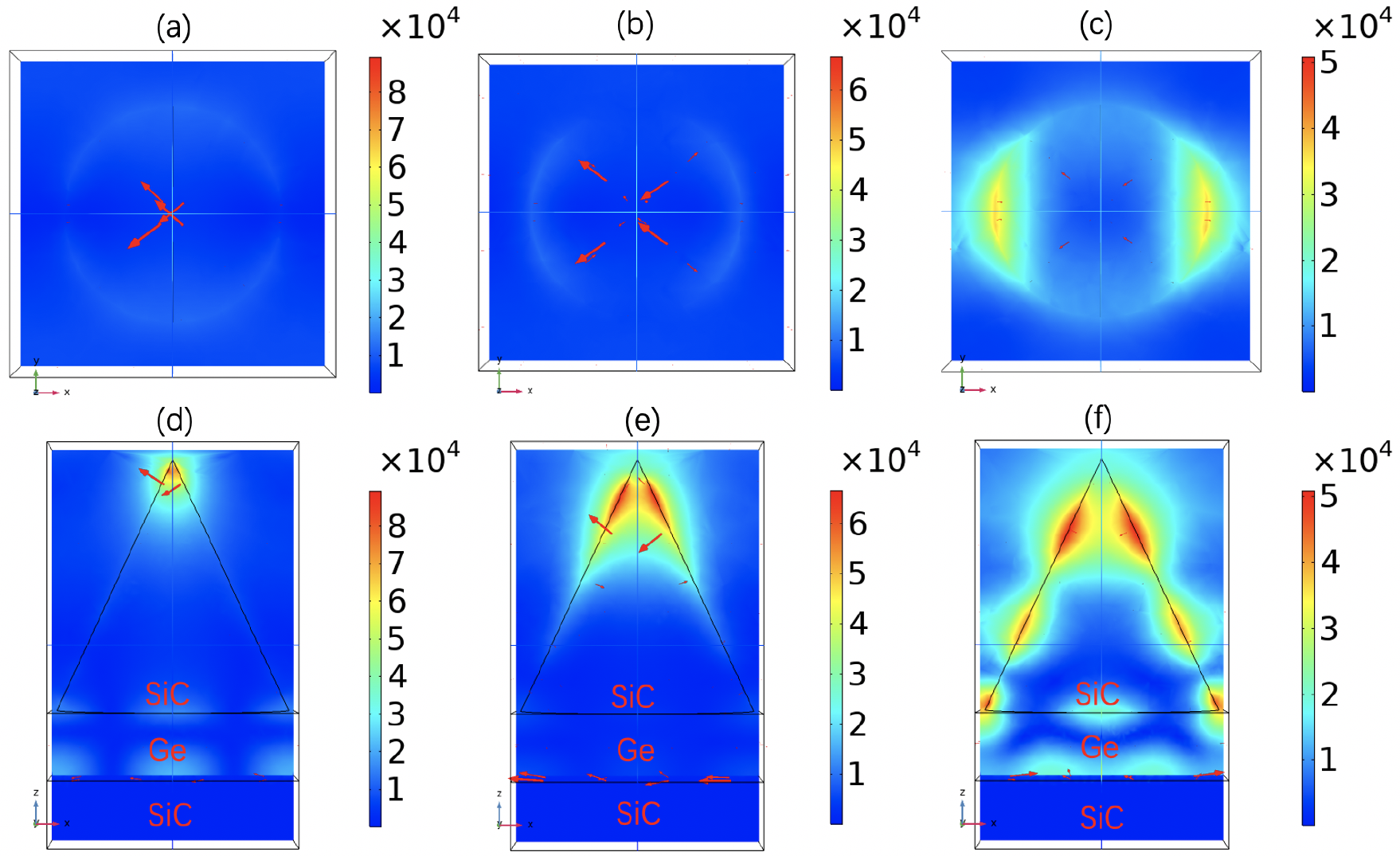}
    \caption{(a)(b)(c) indicate the magnetic field distribution in 10.7, 10.9, 11.2 $\mu m$, respectively and (c)(d)(e) indicate the corresponding magnetic field in xy-plane. The red arrows represent the scale and directions of displacement current density}
    \label{f1}
\end{figure}

 Similarly, we choose 3 typical wavelengths, 10.6, 10.8 and 11.1 $\mu m$, respectively. {The magnetic field distribution is shown in Fig.~\ref{f1}.}
\section{The PLC circuit model}
The PLC model hass successfully employed by many magnetic polaritons model including both metal\cite{lee2008coherent,wang2012wavelength,sakurai2014resonant} and polar dielectric structure like {SiC}\cite{wang2011phonon}.
\begin{figure}[hbt!]
    \centering
    \includegraphics[width=30pc]{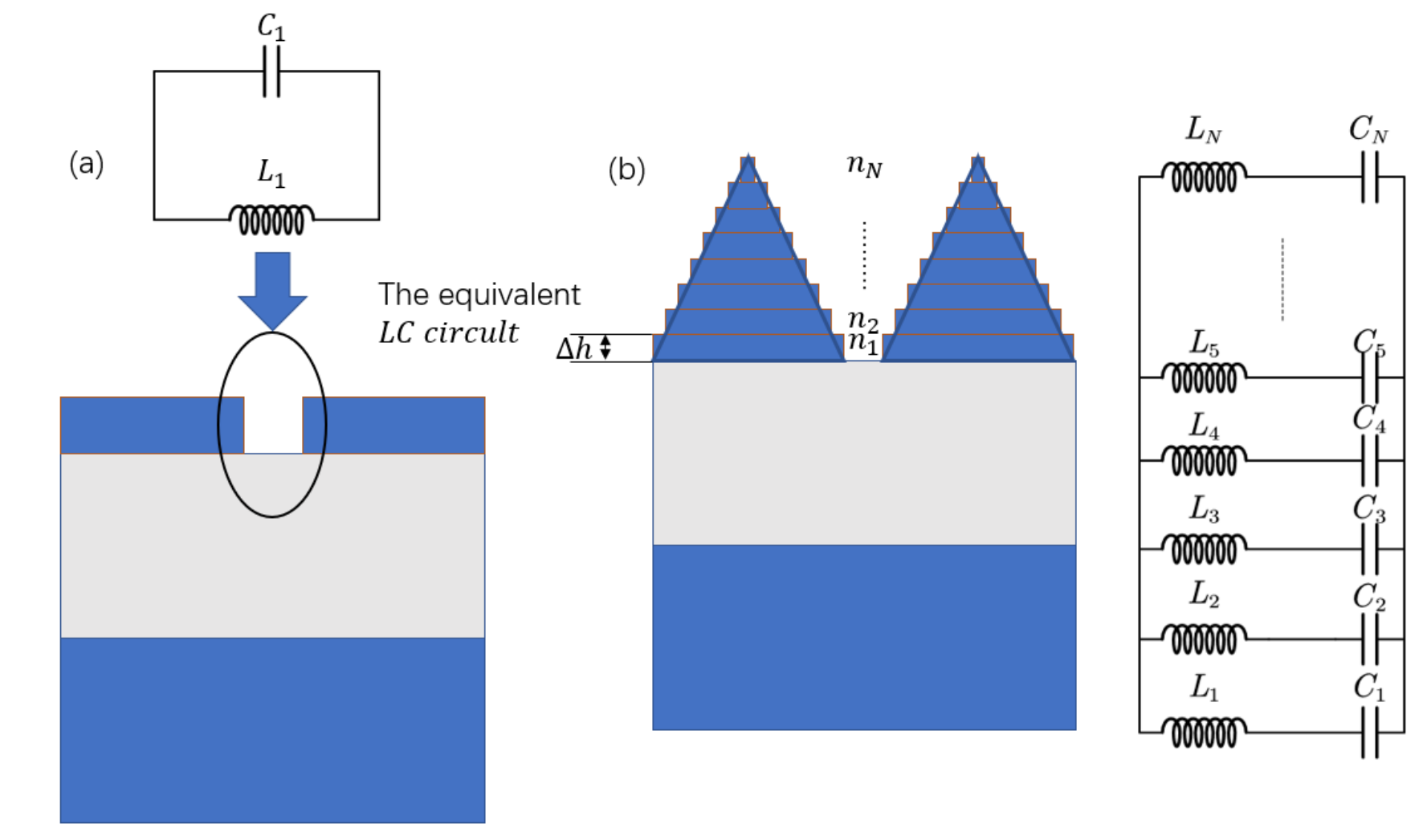}
    \caption{(a)The equivalent LC circuit of square structure. (b)The approximate model of pyramid structure and equivalent PLC circuit}
    \label{PLC}
\end{figure} 
\par Here we present {an} equivalent LC circuit model for the square grating structure in Fig.~\ref{PLC}(a). {It turns out that it can be successfully applied to 1D structure.}
As for the pyramid structure. We can approximately view it as many squares vertically stacking over the substrate with same height $\Delta h$ and its width evenly decrease.
Each of them becomes a distinct resonator and is connected with each other in parallel.
The real situation may not look like this, but this model can {present} us an intuitive demonstration.
\par From basic physics, we have the capacitance model of two metal parallel plate. However, when applying to the polar dielectric material, a numeric factor occurs\cite{zhou2006unifying}. 
\begin{equation}
    C_n = c_1\frac{\varepsilon_0 \Delta h}{(T-\Lambda)+2n \sin{\theta} \Delta h},
\end{equation}
where $\theta$ indicate the bottom corner of the pyramid and $c_1$ accounts for the nonuniform charge distribution at the strip surfaces due to the induced electric currents with a range of $0.2<c<10.3$\cite{zhou2006unifying}.
Two contributions make up the inductance $L_{n}$:
\begin{itemize}
    \item Mutual inductance obtained for each plate $L_{m,n}$
    \item Kinetic inductance that arrised from the kinetic energy of mobile charge carriers $L_{k,n}$
\end{itemize}
It can be written in an uniform way\cite{wang2011phonon,zhou2006unifying,zhang2007nano}:
\begin{equation}
    L_n = L_{m,n}+L_{k,n} = 0.5\mu_0\Delta h b - \frac{\Delta h}{\omega^2 \varepsilon_0 \varepsilon' \delta},
\end{equation}
where $\varepsilon_0$ and $\mu_0$ indicate the vacuum permittivity and permeability, respectively.
Also notice that those formula we bring out above are given in per unit length of our structure since the magnetic polaritons are independent on the structure length along the direction of magnetic field. Therefore, it is proper to present in that way.
The total impedance of $n$th resonator can be written as:
\begin{equation}
    Z_n = i\omega(L_{m,n}+L_{k,n}-\frac{1}{\omega^2 C_1}).
    \label{res}
\end{equation}
The total impedance of pyramid structure is (connected in parallel):
\begin{equation}
    \frac{1}{Z_{total}} = \frac{1}{Z_1} +\frac{1}{Z_2}+\cdots +\frac{1}{Z_n}.
\end{equation} 
\par The resonance frequency can be find by zeros of (\ref{res}). By setting 10.82 $\mu m$ as one of the solutions and $\Delta h = 0.6~\mu m$, the factor $c_1$ can be determined as $c_1 = 0.26$.
Then, 10.33, 10.39, 10.45, 10.58 and 10.71 $\mu m$ can be verified as the solutions of \ref{res} corresponding to different $n$ by $\mathbf{mathematica}$. By applying Lorentz line type to each solution and we have similar line type in Fig.~\ref{fig1}.

\section{The spectral response of different incident angles}
\begin{figure}[hbt!]
    \centering
    \includegraphics[width=30pc]{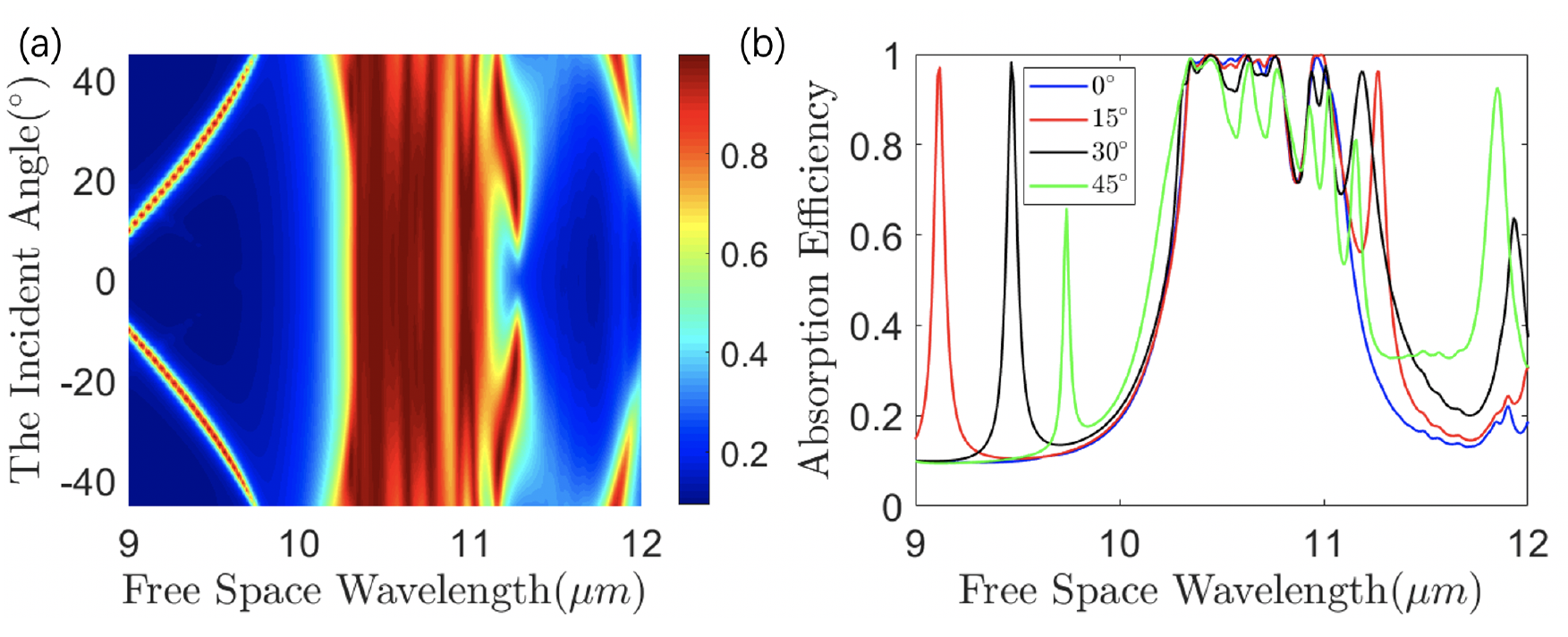}
    \caption{(a)The angle response of incident angle from $-45^{\circ}$ to $45^{\circ}$.(b)The spectral with incident angle 0, 15, 30, 45$^{\circ}$.}
    \label{angle_response}
\end{figure}
Fig.~\ref{angle_response} show the dependence of the spectral response with different incident angle. One may conclude from that the perfect absorption doesn't limited to normal incidence but can be extended to a large angle. Also, an additional peak was excited and shift with the change of incident angle ({basically} linearly).
\par
\begin{figure}[hbt!]
    \centering
    \includegraphics[width=30pc]{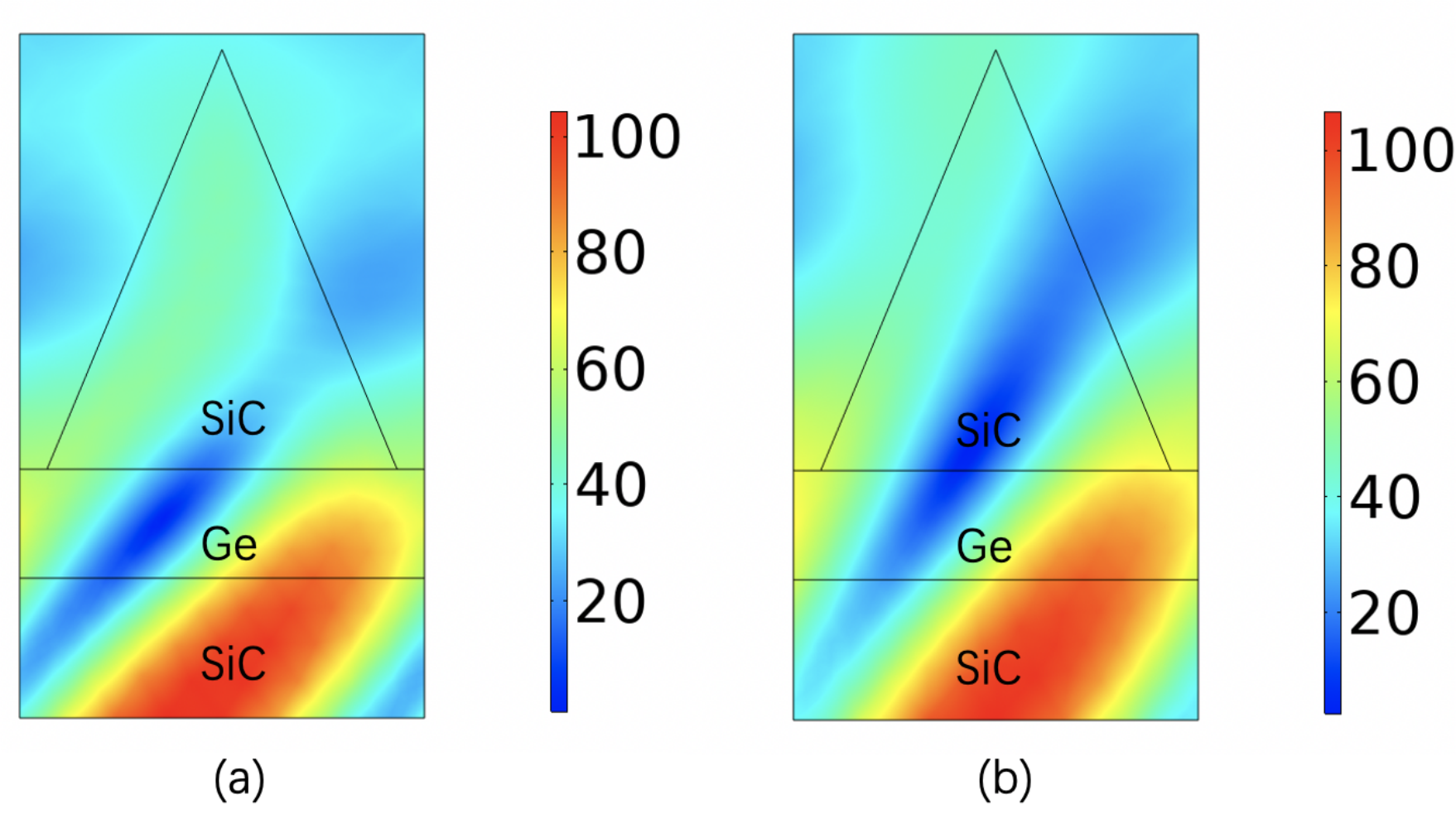}
    \caption{(a)The magnetic field distribution with incident angle 15$^{\circ}$ and wavelength 9.15 $\mu m$(b)The magnetic distribution with incident angle 30$^{\circ}$ and wavelength 9.5$\mu m$. (a) and (b) corresponding to the first peak and second peak in Fig.~\ref{angle_response}, respectively.}
    \label{gmr}
\end{figure}

 In order to investigate the additional peaks in Fig.~\ref{angle_response}, the magnetic field distributions was given in Fig.~\ref{gmr}. One may conclude that the additional peaks originate from the guided-mode resonance.
\section{The geometric effect}
\subsection{Bridge the dip}
By investigating the absorption spectrum of the structure, one {finds} that there is a dip appeared around $10.9~\mu m$ and it may {cause} serious problems when it comes to practical applications. With the discussion above, one is convinced that {the depth of the dip has strong dependence of the filling factor of the grating.} 
By optimizing the structure, the absorption efficiency is over 90\% and one may roughly say that the dip is bridged.
\begin{figure}[hbt!]
    \centering
    \includegraphics[width=30pc]{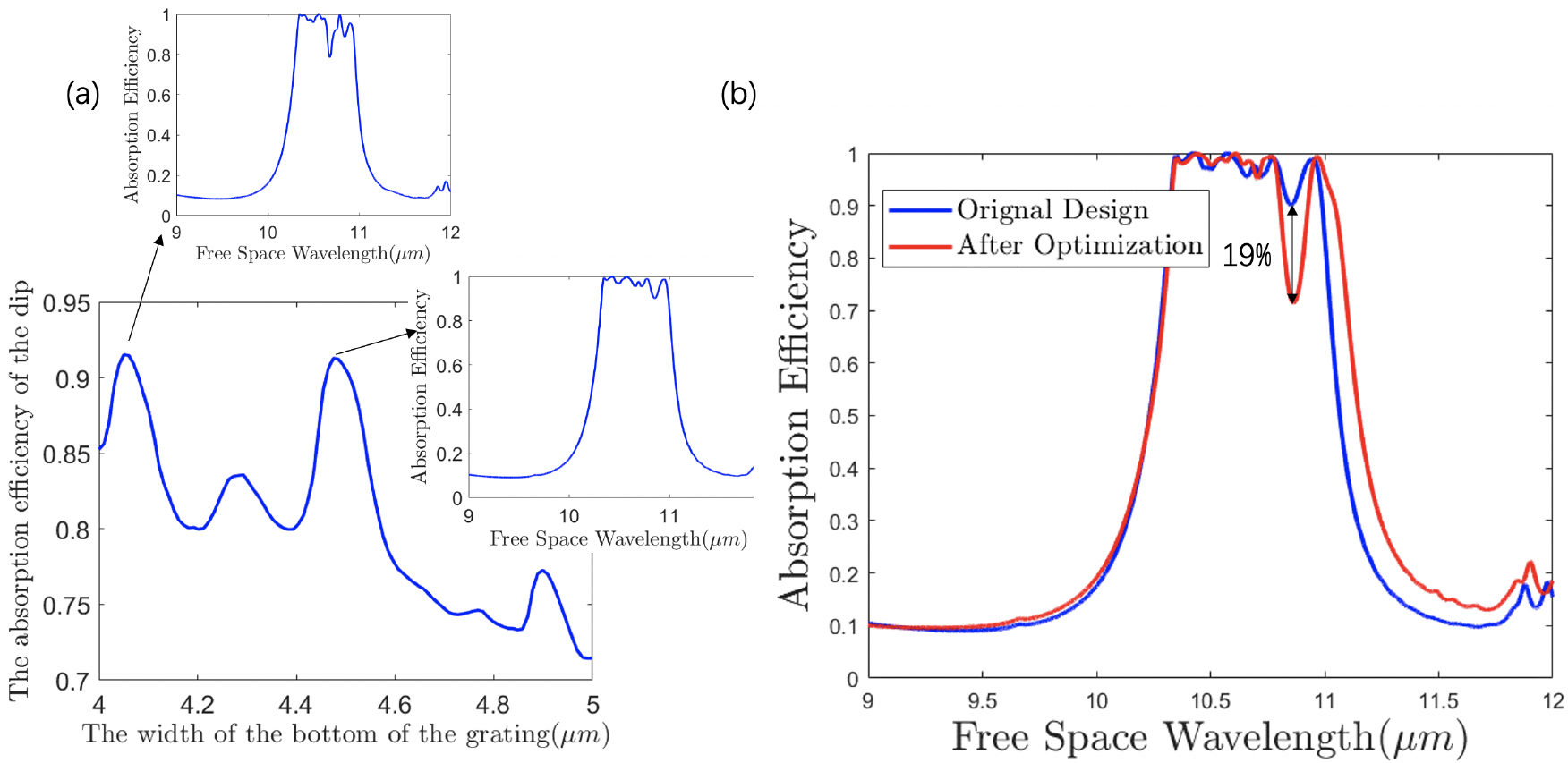}
    \caption{(a)The width of the bottom of the grating versus the absorption efficiency of the dip that occur in 10.86 $\mu m$ and the insets are the shape of the absorption spectrum of the two peaks, respectively. (b)The comparison between the spectrum of the original design and the response after optimization.}
    \label{fig5}
\end{figure}
\par From Fig.~\ref{fig5}, one may vividly find that by choosing the width of the bottom as 4.5 $\mu m$ can fulfil our requirement. As Fig.~\ref{fig5} indicates, the spectrum has similar shape except for the dip. As we demonstrate before, 
{The last peak} in 11 $\mu m$ has different cascades resonance behavior from the others. In this case, by choosing the geometry parameter reasonably, the last peak can also be linked with the others therefore the spectrum is effectively broadened.
\subsection{The trapezoidal structure}
\begin{figure}[hbt!]
    \centering
    \includegraphics[width = 25pc]{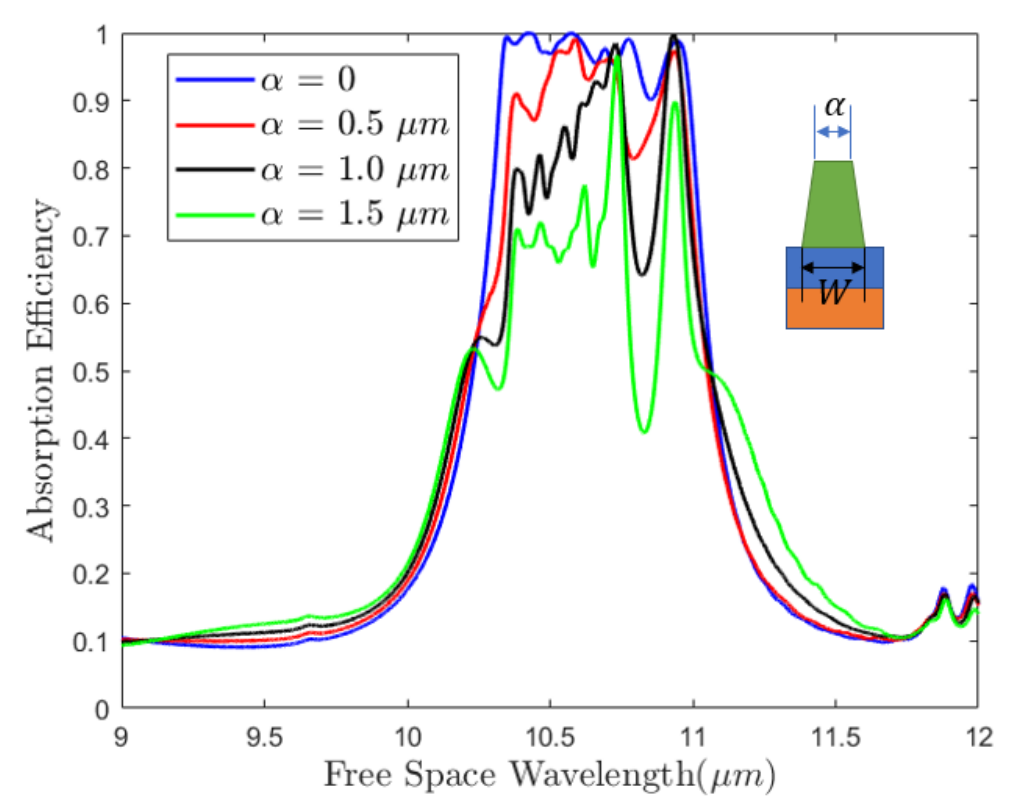}
    \caption{The spectral response with different $\alpha$}
    \label{Fig6}
\end{figure}
Further, we investigate the influence of the width of the top (imagine the grating is knifed in the top and consequently it becomes a trapezoidal). Fig.~\ref{Fig6} demonstrate that {the greater the width of top of the grating, the more the resonance peaks are suppressed}. However, the position of the peak is roughly remain unshifted.
\section{Conclusion}
We  present a pyramid-like polar dielectric material based grating-waveguide structure which exhibit the perfect absorption property in the regime of atmospheric window. The PLC circuit model has been successfully applied to explain the absorption mechanism and help us modify the geometry of the structure. The absorption band is subsequently broadened.
By investigating both the 1-D and 2-D structure, it has been observed that the omnidirectional and polarization insensitive properties can be realized which may exhibit a great potential for meteorology applications or other practical applications such as band-stop filters, detection and imaging at invisible frequencies.
\section*{Acknowledgements}
This research has been funded by National Science Fundation of China (No.41675154). 
\bibliographystyle{elsarticle-num}
\bibliography{cite}
\end{document}